\documentclass[runningheads]{llncs}
\usepackage{graphicx}

\usepackage{tikz}
\usepackage{comment}
\usepackage{amsmath,amssymb} 
\usepackage{color}
\usepackage[intoc]{nomencl} 
\usepackage{amsmath}
\usepackage{amssymb}
\usepackage{booktabs}
\usepackage{epsfig}
\usepackage{color}

\usepackage{enumitem}
\usepackage{IEEEtrantools}
\usepackage{mathtools}
\usepackage{url}
\usepackage{caption}
\usepackage{subcaption}
\usepackage{multirow}
\usepackage{tabularx}
\usepackage{booktabs}
\usepackage{nccmath}

\usepackage[pagebackref,breaklinks,colorlinks]{hyperref}
\usepackage[capitalize]{cleveref}
\crefname{section}{Sec.}{Secs.}
\Crefname{section}{Section}{Sections}
\Crefname{table}{Table}{Tables}
\crefname{table}{Tab.}{Tabs.}
\usepackage[accsupp]{axessibility}  

\begin{document}
\pagestyle{headings}
\mainmatter
\def\ECCVSubNumber{3923}  

\title{SJ-HD$^2$R: Selective Joint High Dynamic Range and Denoising Imaging for Dynamic Scenes} 


\titlerunning{SJ-HD$^2$R: Selective and Joint HDR and Denoising Imaging}
%
\author{Wei Li\inst{1\footnotemark[1]} \and
Shuai Xiao\inst{1\footnotemark[1]} \and
Tianhong Dai\inst{2}
Shanxin Yuan \inst{2} \and
Tao Wang \inst{1} \and
Cheng Li \inst{1} \and
Fenglong Song \inst{1}
}
\authorrunning{Wei Li, Shuai Xiao, Tianhong Dai, Shanxin Yuan et al.}
%
\institute{Huawei Noah's Ark Lab \and Imperial College London}
\renewcommand{\thefootnote}{\fnsymbol{footnote}}
\footnotetext[1] {These authors contributed equally to this work.}

\maketitle
\begin{abstract}

Ghosting artifacts, motion blur, and low fidelity in highlight are the main challenges in High Dynamic Range (HDR) imaging from multiple Low Dynamic Range (LDR) images. These issues come from using the medium-exposed image as the reference frame in previous methods. To deal with them, we propose to use the under-exposed image as the reference to avoid these issues. However, the heavy noise in dark regions of the under-exposed image becomes a new problem. Therefore, we propose a joint HDR and denoising pipeline, containing two sub-networks: (i) a pre-denoising network (PreDNNet) to adaptively denoise input LDRs by exploiting exposure priors; (ii) a pyramid cascading fusion network (PCFNet), introducing an attention mechanism and cascading structure in a multi-scale manner. To further leverage these two paradigms, we propose a selective and joint HDR and denoising (SJ-HD$^2$R) imaging framework, utilizing scenario-specific priors to conduct the path selection with an accuracy of more than 93.3$\%$. We create the first joint HDR and denoising benchmark dataset, which contains a variety of challenging HDR and denoising scenes and supports the switching of the reference image. Extensive experiment results show that our method achieves superior performance than previous methods.

\keywords{high dynamic range, denoising, computational photography, deep learning}
\end{abstract}
\begin{figure}[t]
	\centering
	\includegraphics[width=1.0\columnwidth]{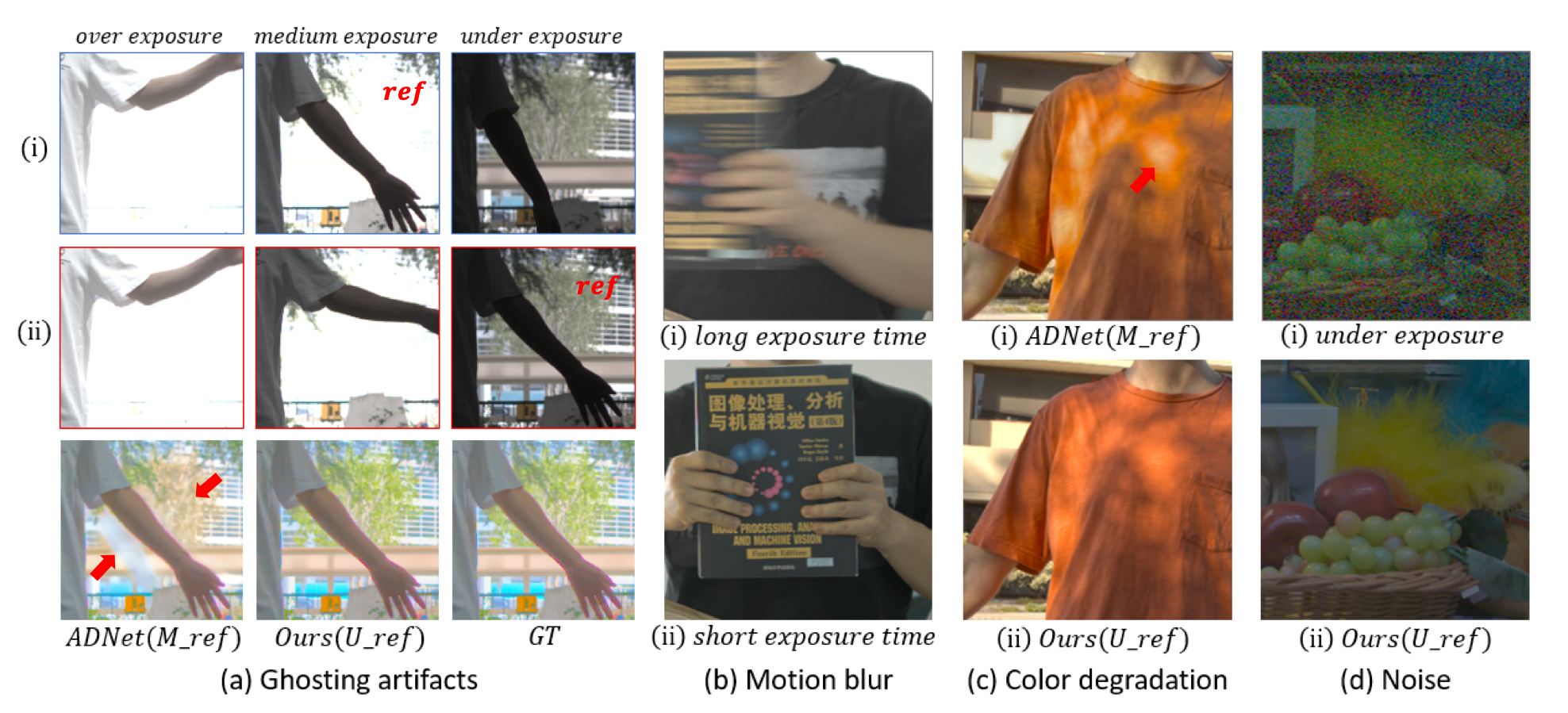}
	\vspace{-4mm}
	\caption{Existing challenges in HDR fusion, $M_{ref}$ and $U_{ref}$ are medium-exposure based and under-exposure based methods respectively. (a) ghosting artifacts caused by occlusion in highlight. (i) and (ii) represent the input LDRs for ADNet~\cite{ADNet} and Ours; (b) motion blur; (c) color degradation; (d) heavy noisy in the dark, the luminance of which is near-zero but is brightened manually;}
	\label{fig:demo}
\end{figure}
\section{Introduction}
HDR imaging is in high demand across numerous industrial applications, such as photographic devices and safety-critical decision-making process in autonomous driving systems \cite{robidoux2021hdr_isp_opt}. However, the dynamic range of typical real-world scenes varies across orders of magnitude ($10^{6}$:1, 120+dB) \cite{radonjic2011dynamic}, which is far beyond the dynamic range that most consumer-grade image sensors can capture in a single shot \cite{debevec2008recovering,eilertsen2017hdr}. Consequently, over-exposed and/or under-exposed regions appear in LDR images. One feasible solution is to capture several LDR images with different exposure times and merge them into an HDR image \cite{debevec2008recovering,mannbeing,granados2010optimal,reinhard2010high,yan2017high}.

While great success has been achieved in this line of research, challenges still exist, especially in dealing with the motion among sequentially-multiplexed LDR images of dynamic scenes. 
The global motion can be well addressed through homographies~\cite{szpak2014sampson,szpak2015robust}. However, the foreground motion is complex and difficult to handle.
Generally, the medium-exposed LDR image is endowed as the reference image to maintain structural consistency with the reconstructed HDR image \cite{sen2012robust,hu2013hdr,kalantari2017deep,kalantari2019deep,wu2018deep,prabhakar2019fast,yan2020deep,yan2019attention,dai2021waveletbased,HDR_GAN_2021,ADNet}. This approach works well in many cases, but it leads to ghosting artifacts, motion blur, and low fidelity, especially when the ghosting artifacts caused by occlusion occurs in highlights\cite{yan2019attention,HDR_GAN_2021}. 
In this case, the reference frame in highlight has already been over-exposed, but the complementary contents expected to merged from supporting frames are occluded, as shown in Figure~\ref{fig:demo} (a), which is an ill-posed issue to hallucinate the missing details. Additionally, the longer exposure time of medium or over-exposed LDRs results in motion blur effect\cite{Mustaniemi2018LSD2J} \textit{even in the LDR images} when there is fast camera motion or object motion, as shown in Figure~\ref{fig:demo} (b). Moreover, medium-exposure based fusion paradigm, as shown in Figure~\ref{fig:demo} (c), empirically causes color degradation in highlight.

The under-exposed LDR image with a shorter exposure time, on the other hand, will produce sharper but noisier image content. Hence, a straight-forward but nontrivial thought is to use the short exposure as the reference, which is the key to the aforementioned issues: ghosting artifacts, motion blur effect, and color degradation. Despite it outperforms the medium-exposed counterpart by large margins, it brings new challenges in dark region, where the under-exposed image may be corrupted by heavy noise (Figure~\ref{fig:demo} (d)), making deghosting and denoising in the dark region difficult. To alleviate the noise in input LDRs, we propose a pre-denoising network (PreDNNet) by introducing $ISO$ prior information to boost denoising performance and control the strength of denoising~\cite{Abdelhamed_2018_CVPR,plotz2017benchmarking,Chen_2018_CVPR}.
 
To further deal with the HDR fusion and deghosting after pre-denoising stage, we propose a pyramid cascading fusion network (PCFNet), exploiting a cascading structure of encode-attention-fusion in multi-scales. In addition, we observe that the extremely wide luminance distribution of an HDR scene makes denoising and deghosting in dark region difficult. To address this deficiency, we propose to train the entire system in tone-mapped domain from the perspective of distribution adaptation.

\begin{figure*}[t]
    \centering
    \includegraphics[width=1.0\textwidth]{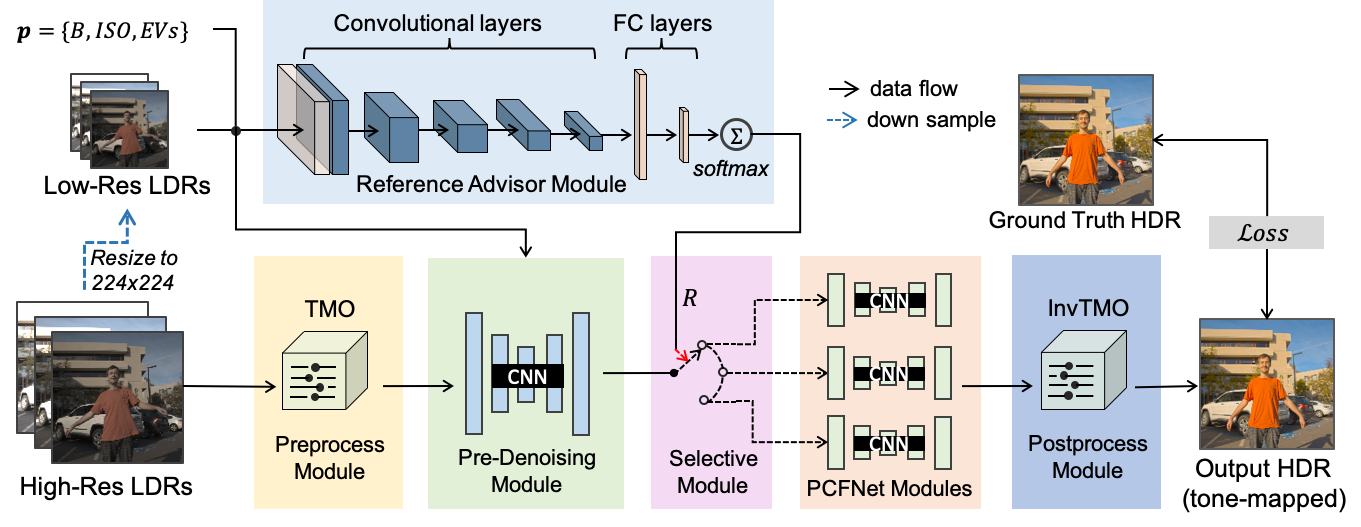}
    \caption{\textbf{SJ-HD$^2$R}: proposed selective and joint HDR and denoising framework. The framework includes two branches: the upper branch named reference advisor (RA) module that takes low-resolution LDRs and scenario-specific priors, e.g., brightness($B$), $ISO$, exposure setting($EV$) as input and outputs parameter $R$; The bottom branch is the joint HDR and denoising imaging pipeline that takes high-resolution LDRs as input and is firstly processed by TMO, pre-denoising sequentially, followed by path selection module triggered by $R$. Multiple PCFNet models using different image as reference are pre-trained and loaded dynamically during inference. The final InvTMO module transform the predicted image back to linear domain for further visualization by adopting editable tone-mapping.}
    \label{fig:overall_structure}
    \vspace{-3mm}
\end{figure*}

A satisfactory compromise between using under or medium exposed LDR as reference does not always exist. From our observations, however, we found that in a specific HDR scenario with given environment brightness, $ISO$, exposure ratios, and luminance distribution from LDR images themselves, we can predict which paradigm will produce better qualitative results. This triggers us to propose our final solution: selective and joint HDR and denoising imaging (SJ-HD$^2$R) framework, as shown in Figure~\ref{fig:overall_structure}, proposing an extra reference advisor network (RANet) that utilizes scenario-specific priors to predict the winner module and execute the path selection during inference.

The main contributions of this work can be summarized as follows: 
\textbf{(i)} We propose a novel HDR and denoising pipeline, and develop multiple models using different exposure as reference, each of which outperforms all previous baselines on multiple datasets. 
\textbf{(ii)} we pave a new way to the community by proposing a novel selective and joint HDR and denoising imaging framework summarized as SJ-HD$^2$R that can make a good combination of multiple models. \textbf{(iii)} We propose the first SJ-HD$^2$R imaging benchmark dataset, containing a variety of challenging HDR and denoising scenes both in RAW and RGB domain, it firstly supports the switching of references. The whole dataset will be made public.

\begin{figure*}[t]
    \centering
    \includegraphics[width=0.95\textwidth]{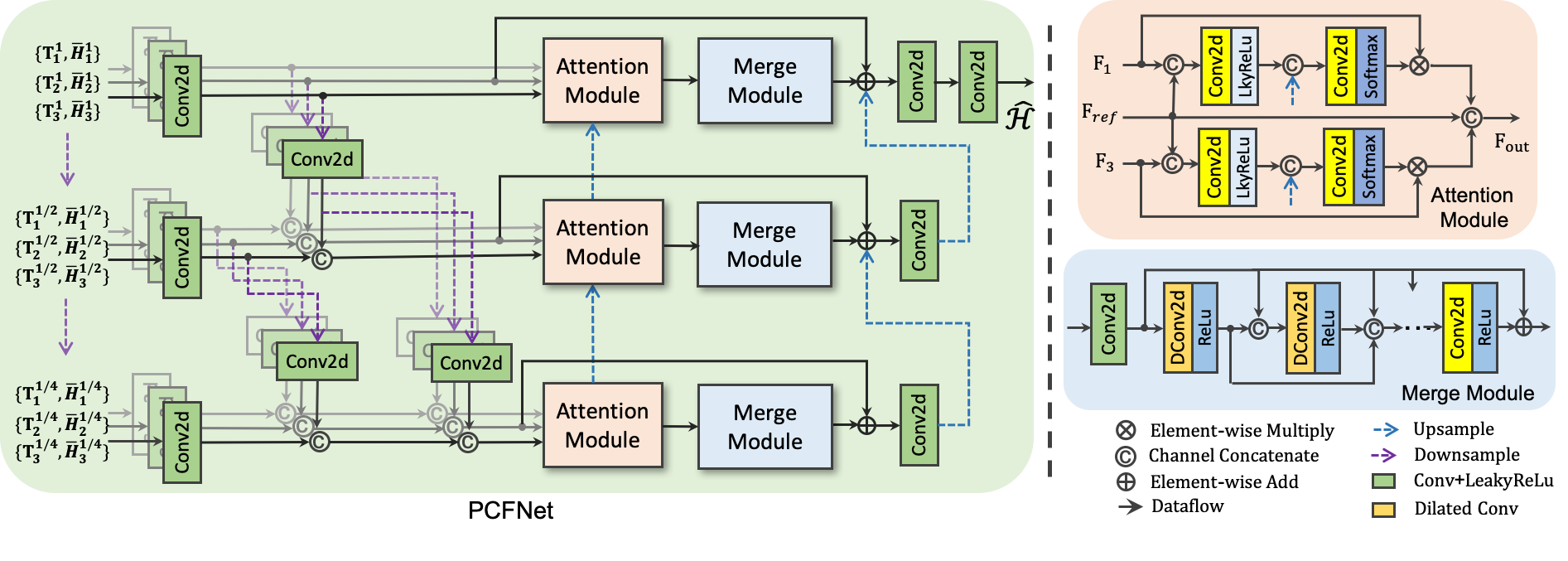}
    \vspace{-3mm}
    \caption{Proposed pyramid cascading fusion network. We design our network with a pyramid paradigm, exploiting a cascading structure of encode-attention-fusion in multi-scales to address problems caused by large motion.}
    \label{fig:pcfnet}
    \vspace{-4mm}
\end{figure*}

\section{Related Work}

In this section, we briefly summarize the HDR datasets and related works on HDR fusion from multi-exposures, as well as burst denoising methods for dynamic scenes.

\textbf{\emph{HDR datasets.}} Dataset plays a key role in deep-learning-based HDR imaging methods. Multiple datasets have been proposed, driving the development of this line of research.  
Sen \emph{et al.}\cite{sen2012robust} and Tursun \emph{et al.} \cite{tursun2016objective} proposed an HDR dataset providing only 8 and 16 scenes without ground truth HDR image. 
\cite{lee2018deep,cai2018learning} provided a synthetic dataset of static scenes with different exposure settings.
Kalantari \emph{et al.}\cite{kalantari2017deep} first collected 74 and 15 pairs for training and evaluation. 
Prabhakar \emph{et al.}\cite{prabhakar2019fast} collected a larger dataset with 582 paired samples following the approach in~\cite{kalantari2017deep}. \cite{Hu_2020_CVPR_Workshops} proposed a synthetic dataset that was produced by game engine. The latest NTIRE2021 challenge in multiple frames HDR fusion track released a new HDR dataset~\cite{perez2021ntire} containing more than 1400+ pairs with synthetic noise.
All these datasets only support using medium-exposed LDR as the reference in RGB domain, which limits the exploration of switching of references or HDR fusion from multi-exposures in RAW domain.

\textbf{\emph{HDR imaging from multi-exposures.}} Traditional HDR fusion methods \cite{debevec1997,khan2006ghost,grosch2006fast,jacobs2008automatic,pece2010bitmap,heo2010ghost,sen2012robust,hu2013hdr,li2014selectively} for dynamic scenes either perform a global alignment on LDR images and then reject misaligned pixels, or search for the best matching patches from supporting frames to reconstruct HDR image. 
Grosch \emph{et al.}\cite{grosch2006fast} rejected misaligned pixels by calculating an error map according to color difference.
Heo \emph{et al.}\cite{heo2010ghost} used joint probability densities to detect motion regions.
Sen \emph{et al.}\cite{sen2012robust} proposed a novel patch-based energy-minimization formulation by integrating alignment and reconstruction in a joint optimization. 
Hu \emph{et al.}\cite{hu2013hdr} proposed to use brightness and gradient consistencies to conduct alignment in the latent transformation domain. 
These methods cannot handle noise well and often lead to unsatisfactory artifacts in reconstructed HDR image.\par

Many deep-learning-based methods~\cite{ram2017deepfuse,kalantari2017deep,prabhakar2019fast,wu2018deep,yan2019attention,yan2019multi,yan2020deep,dai2021waveletbased,HDR_GAN_2021,ADNet,Chen_2021_ICCV} have been proposed and show great success.
Despite several unsupervised methods~\cite{Prabhakar_2021_CVPR,ram2017deepfuse} were proposed and achieved good performance without labeled data, they still failed in some extreme scenes.
Model-based methods either explicitly align the LDRs with optical flow or implicitly implement feature alignment by introducing attention mechanism.
Kalantari \emph{et al.}~\cite{kalantari2017deep} and Prabhakar \emph{et al.}~\cite{prabhakar2019fast} first calculated the optical flow to align the supporting LDRs to the reference image, and then fused the aligned stack of LDRs using convolutional neural network (CNN). However, the estimation error of optical flow, usually happens in extreme-light condition (e.g., over-exposed regions or under-exposed regions with heavy noise), will propagate and be accumulated to following fusion network, leading to artifacts in reconstructed HDR image.
Wu \emph{et al.}~\cite{wu2018deep} performed global registration on LDR images, and used a U-Net~\cite{Unet} to learn the mapping from LDR images to HDR image. 
Yan \emph{et al.}~\cite{yan2019attention} proposed a deep CNN by introducing attention modules and dilated residual dense blocks to deal with large motion. 
Niu \emph{et al.}~\cite{HDR_GAN_2021} first introduced GAN~\cite{NIPS2014_5ca3e9b1} into HDR reconstruction, contributing to faithful information in over-exposed regions. 
Dai \textit{et al.}~\cite{dai2021waveletbased} proposed a wavelet-based network which dealt with ghosting artifacts in different frequency sub-bands. 
Liu \emph{et al.}~\cite{ADNet} processed LDR image with a PCD alignment module~\cite{wang2019edvr}, and a spatial attention module\cite{yan2019attention} separately, achieving state-of-the-art (SOTA) performance in NTIRE2021 challenge~\cite{perez2021ntire}. All above methods use medium-exposure as reference and suffer from the ill-posed information loss issue in saturated regions. While Prabhakar \emph{et al.}~\cite{prabhakar2019fast} discussed about the influence of choice of reference image on ghosting artifacts, they do not give any in-depth analysis. As a result, all these methods produce severe ghosting artifacts and color degradation in saturated regions.

As we first emphasize the importance of utilizing under-exposed image as the reference for addressing several existing issues, it is necessary to compared this paradigm with burst denoising methods~\cite{Mildenhall_2018_CVPR,Godard_2018_ECCV,TIAN2020251,10.1145/2980179.2980254} or denoising with high exposure images~\cite{Mustaniemi2018LSD2J,7953319,9127139}. Generally, burst denoising methods have small motion and acquire intermediate SNR but need more than 8 under-exposed images~\cite{10.1145/1143844.1143845}, which is expensive on time and power consumption and is infeasible for HDR video imaging. High exposure based denosing or debluring methods~\cite{Mustaniemi2018LSD2J} is somehow similar with our work but they failed to handle large object motion.

\section{Methodology}
\subsection{Formulations}
Given a set of LDR images ${[I_1, I_2, ..., I_n]}$ with different exposure times, the aim of HDR imaging is to reconstruct an HDR image that has rich details in both over-exposed and under-exposed regions. 
Generally, the HDR image should be structurally consistent with the reference image ${I_{ref}}$. In this paper, we follow previous works~\cite{kalantari2017deep,wu2018deep,yan2019attention} that take three LDR images as input. Instead of using the medium-exposed LDR image as the reference, we also consider the under-exposure or over-exposure LDR image as the reference. Moreover, we explore HDR imaging in both RGB and RAW domain. 
We first convert LDR images to HDR domain ${H_i, i=1,2,3}$ by a gamma correction and exposure alignment:
\begin{equation}
    H_{i} = {I_{i}^{\gamma}}/t_{i}, i=1,2,3
\label{eq:gamma_corr}
\end{equation}
where $\gamma$ denotes the parameter of gamma correction. We set $\gamma$ to 2.2 for RGB domain image, and 1.0 for RAW domain, as the RAW images are already in linear domain. We normalize $I_{i}$ from linear domain ($L_i=I_{i}^{\gamma}$) into HDR domain ${H_i}$ to align the luminance among LDRs with exposure time $t_{i}$.
We use ${N^{\mathcal{R}}}$, ${N^{\mathcal{D}}}$, and ${N^{\mathcal{P}}}$ to represent our RANet, PreDNNet and PCFNet respectively, so that our SJ-HD$^2$R imaging framework can be formulated as follow:
\begin{equation}
    {\mathcal{R}} = N^{\mathcal{R}}(\emph{\textbf{p}}, I_{ls}, \theta^\mathcal{R})
\label{eq:RANet}
\end{equation}
where \emph{$\textbf{p}$} and $I_{ls}$ represent the scene priors and low resolution LDRs. $\theta^{\mathcal{R}}$ represents the learnable weights of ${N^{\mathcal{R}}}$. Here $\mathcal{R} \in \{0, 1, 2\}$, representing the classification results for three models.
\begin{equation}
    {\widehat{H}_T} = N^{\mathcal{P}}([N^{\mathcal{D}}(\mathcal{T}([L, ISO]), {\theta}^{\mathcal{D}}), \mathcal{T}(H)], {\theta}^\mathcal{P}, \mathcal{R})
\label{eq:JHDI}
\end{equation}
where ${\widehat{H}_T}$ is the output image in non-linear domain. $L$ and $H$ denote input LDRs in linear and HDR domain respectively. ${\theta}^\mathcal{D}$ and ${\theta}^\mathcal{P}$ represent the parameters of ${N^{\mathcal{D}}}$ and ${N^{\mathcal{P}}}$, $[\cdot,\cdot]$ denotes the concatenation operator, $ISO$ is a prior metadata provided by our benchmark dataset. $\mathcal{T}$ represents the brightness adjustment function, here we use $\mu$-law tone-mapping function:  

\begin{equation}
    \mathcal{T}(\mathcal{X}) = {\log(1+\mu \mathcal{X})}/{\log(1+\mu)}
\label{eq:TMO}
\end{equation}
where ${\mathcal{X}}$ is the linear input.
$\mu$ is set to 5000 as default. For evaluation and display, we transform $\widehat{H}_T$ back to linear domain by its inverse function:
\begin{equation}
    {\mathcal{V}(\mathcal{X})} = {((1+\mu)^\mathcal{X} - 1) / \mu}
\label{eq:invTMO}
\end{equation}
where $\mu$ remains consistent with the setting in \eqref{eq:TMO}.




\par

\subsection{Overview of Our Framework}
The proposed SJ-HD$^2$R framework includes two branches as shown in Figure~\ref{fig:overall_structure}: (i) the reference advisor branch that takes the low-resolution LDRs and scenario-specific priors, e.g., brightness ($B$), $ISO$, exposure setting ($EV$) as input and outputs parameter $R$, encoding the preferred path selection; (ii) the second branch is our joint HDR and denoising imaging pipeline that takes high-resolution LDRs as input and is firstly processed by TMO, pre-denoising sequentially, followed by path selection module triggered by $R$. Multiple PCFNet models using different exposure as reference are pre-trained and will be load dynamically during inference. The details of our PCFNet is illustrated in Figure~\ref{fig:pcfnet}. The final InvTMO module transform the predicted image back to linear domain for further visualization by adopting editable tone-mapping operations. The output HDR image and the ground truth HDR are best viewed by \emph{Photoshop Software}~\cite{photoshop}.
\subsection{Reference Advisor Network} 
We use a compact CNN to predict the ground truth of reference image, named RANet, taking as input the low resolution LDRs resized to 224x224 and available scene priors, e.g., brightness ($B$), $ISO$, exposure setting ($EV$) information and formulating it to a simple image classification problem. We use the tone-mapped HDR-VDP metrics as the classification label in our experiment setup, as it best indicates the visual quality of HDR scenes, and other metrics, e.g., PSNR or SSIM are biased either in highlight or in the dark. Experiments show its powerful performance (achieving 93.3$\%+$ accuracy), with very limited additional computational consumption. Specifically, our RANet costs only 2.05 \emph{Gmacs} and 270 $K$ parameters. Architecture detail of RANet can be found in supplement.

\subsection{Joint HDR and Denoising Pipeline}
Our joint HDR fusion and denoising pipeline is constituted of reversible and differential tone-mapping modules, pre-denoising network, and pyramid cascading network that are detailed as below:

\textbf{\emph{Tone mapping operation (TMO).}} To balance the ultra-wide luminance distribution of the real world, we propose to transform the given LDRs to a non-linear domain and its inverse by \emph{Eq}.\ref{eq:TMO} and \emph{Eq}.\ref{eq:invTMO}. Experiment results proved that this operation leads to a significant boost on restored image quality.

\textbf{\emph{Pre-denosing network.}} To alleviate heavy noise in dark region, we propose the PreDNNet to denoise the input LDRs before merging. Our proposed PreDNNet use a U-Net~\cite{Unet} as backbone and is trained with existing HDR paired dataset, which means there is no need to build up an extra denoising dataset for training and evaluation. 
We use non-linear LDRs and ground truth ${[T_{s}^{i}, \overline{H}_{s}^{i}], i=1,2,3}$ to demonstrate the supervised training pairs in non-linear domain, where the ground truth for LDR image ${T_{s}^{i}={\mathcal{T}}(L_{s}^{i})}$ is ${\overline{H}_{s}^{i}={\mathcal{T}}(\overline{H}*t_{i})}$. So that the proposed PreDNNet can be formulated as:
\begin{equation}
    \widehat{T}_{s}^{i} = {N^{\mathcal{D}}}(T_{s}^{i}, ISO, {\theta}^{\mathcal{D}}), i=1,2,3
\label{eq:PreDNNet}
\end{equation}
where ${N^{\mathcal{D}}}$ and ${\theta}^{\mathcal{D}}$ represents the PreDNNet and its weights respectively. 
$\widehat{T}_{s}^{i}$ is the predicted denoised images of $N^{\mathcal{D}}$.


\textbf{\emph{Pyramid cascading fusion network.}} The PCFNet, as shown in Figure~\ref{fig:pcfnet}, takes the output of PreDNNet as input and generates a merged HDR image. In PCFNet, we design our network with a pyramid paradigm, which can effectively broaden the receptive field and thus be capable of compensating large motion. We integrated several effective modules from previous methods. We summarize the main features of our PCFNet as follow: 
 \subitem{\textbf{\emph{Pyramid encoder}} extracts features from input LDRs at multi-scales. Different with previous encoding approaches in \cite{Unet,wu2018deep,yan2019attention,dai2021waveletbased}, we extract hierarchical feature not only in feature space but also in images space~\cite{Trinidad2019MultiViewIF}. In addition, this type of top-down downsampling can provide coarse-to-fine information for deeper-scale network. All down-sampling operators use $\times2$ average pooling.}\par


\subitem{\textbf{\emph{Multi-scale attention module (MSA)}}
adopts attention mechanism into multi-scale to alleviate the ghosting problem, extracting complementary information from supporting frames, and it rationally combines the temporal attention among multi-exposures with spatial attention in multi-scales. In each attention module in Figure \ref{fig:pcfnet}, it takes as input encoded feature of multi-exposures ${[F_{1}, F_{ref}, F_{3}]}$, in which ${F_{ref}}$ is the feature map of reference, then a convolution layer is applied to the concatenated feature maps ${[F_{ref}, F_{i}], i=1,3}$ followed by a $sigmoid$ function. The attention maps are then applied to the corresponding neighbour frame by element-wise multiplication.}\par

\subitem{\textbf{\emph{Pyramid merging and reconstruction}} {conducts the merging and reconstruction at the end stage of PCFNet. We introduce dilated residual dense block (DRDB) \cite{yan2019attention} for merging and reconstruction. More detail about DRDB can be found in \cite{yan2019attention}. Different from \cite{yan2019attention} that uses three DRDBs for reconstruction, we only need one DRDB but in each pyramid scale which obtains a larger receptive field. At the tail of this module, a residual addition is applied on feature space, adding reference feature and the feature of deeper scale by up-sampling. In this work we use $\times2$ bilinear interpolation.}}\par



\subsection{Training Loss}
Two types of loss functions are used to train our networks: reconstruction loss $\mathcal{L}_{\mathcal{R}}$ and \emph{sobel} loss ${\mathcal{L}_{sobel}}$. We adopt $\ell_{1}$ loss as the reconstruction loss, as it has been proved effective in image restoration tasks \cite{yan2019attention}. 
We introduce ${\mathcal{L}_{sobel}}$ in \cite{dai2021waveletbased} to further boost performance for HDR reconstruction. Different from previous work~\cite{kalantari2017deep,yan2019attention,wu2018deep} that use tone-mapped $\ell_{1}$ loss , the output image of our model is already in tone-mapped domain.
The reconstruction loss is defined as below:
\vspace{-2mm}
\begin{equation}
    \mathcal{L}_{\mathcal{R}} = ||{\hat{H}_{T} - \mathcal{T}(\overline{H})}||_{1}
    \vspace{-2mm}
\end{equation}

where $\hat{H}$ is the predicted image of $N^{\mathcal{P}}$ and $\overline{H}$ is the ground truth.
In order to restore the structure information in the generated HDR image, we also use the \emph{sobel} loss, which is defined as:
\begin{equation}
    \mathcal{L}_{sobel} = ||{\nabla_{x}\hat{H}_T - \nabla_{x}\mathcal{T}(\overline{H})}||_{1} + ||{\nabla_{y}\hat{H}_T - \nabla_{y}\mathcal{T}(\overline{H})}||_{1}
\label{eq:sobel_loss}
\end{equation}

where $\nabla_{x}$ and $\nabla_{y}$ are the $sobel$ edge operators in the $x$ direction and $y$ direction of the image, respectively. Our final loss is defined as:
\begin{equation}
    \mathcal{L}_{total} = (1-\lambda)\cdot\mathcal{L}_{\mathcal{R}} + \lambda\cdot \mathcal{L}_{sobel}
\end{equation}
$\lambda$ is a balancing parameter and set to 0.5 as default.

\subsection{Implementation Details}
We implement our framework using Pytorch and train it on $4\times$ V100 NVIDIA GPUs with 32GB memory. During training, Adam~\cite{kingma2014adam} is selected as the optimizer. We train the PreDNNet and PCFNet for 10000 epochs with batch size 8 and set the initial learning rate to $5\times10^{-4}$ and set the decay with ${gamma=0.7}$ after each 200 epochs until it drops to ${10^{-6}}$. To augment the training data, we randomly crop the LDR images with patch size $256\times256$ and apply random rotation and flipping on the cropped patches. Similarly, we train our RANet for 200 epochs with batch size 16 and initial learning rate 1e-4 using $cosine$ decay strategy. We resize the input LDRs to $224\times224$ resolution to extract global image feature. The whole training takes around 5 days. 

\section{Experiments}
\begin{table*}[!t]
	\centering
	 \resizebox{0.87\textwidth}{!}{
	 \vspace{-5mm}
	\begin{tabularx}{\textwidth}{p{0.14\textwidth}c c c c c c c c c c c}
		\toprule[1pt]
		\multirow{2}{100em}{Method} & \multicolumn{5}{c}{RGB-HDR Dataset} & \multicolumn{5}{c}{RAW-HDR Dataset} \\
		\cmidrule(lr){2-6} \cmidrule(lr){7-11} \cmidrule(lr){8-11}
		&  \scriptsize{{P}-$\mu\uparrow$} & \scriptsize{{P}-{$L\uparrow$}} & \scriptsize{{S}-$\mu\uparrow$} & \scriptsize{{S}-{$L\uparrow$}} & \scriptsize{{VDP3$ \uparrow$}} & \scriptsize{{P}-$\mu\uparrow$} & \scriptsize{{P}-{$L\uparrow$}} & \scriptsize{{S}-$\mu\uparrow$}  & \scriptsize{{S}-{$L\uparrow$}}  & \scriptsize{{VDP3}$ \uparrow$}\\ 
		\hline
		
		Sen\cite{sen2012robust} & 37.50 & 38.39 & 0.9830 & 0.9926 & 8.999 & 36.25 & 37.77 &  {\color{black} \underline{0.9851}} & {\color{black} \underline{0.9922}} & 9.821 \\
		
		Hu\cite{hu2013hdr}  & 33.97 & 31.03 & 0.9729 & 0.9797 & 9.329 & 32.49 & 29.92 & 0.9733 & 0.9728 & 9.205 \\
		
		Kalantari\cite{kalantari2017deep}  & 38.13 & 40.51 & {\color{black} \underline{0.9866}} & {\color{black} \underline{0.995}} & 9.871 & - & - & - & - & - \\
		
		Wu\cite{wu2018deep}  & 38.12 & 37.48 & 0.9725 & 0.9857 & 9.789 & 38.67 & 38.26 & 0.9709 & 0.9897 & 9.854 \\
		
		SCHDR\cite{prabhakar2019fast}  & 39.75 & {\color{black} \underline{40.85}} & 0.9783 & 0.9124 & {\color{black} \underline{9.908}} & - & - & - & - & - \\
		
		AHDR\cite{yan2019attention}  & 38.76 & 38.73 & 0.9772 & 0.9887 & 9.880 & 39.04 & 39.04 & 0.9757 & 0.9915 & 9.863 \\
		
		DWT\cite{dai2021waveletbased} & 39.09 & 38.30 & 0.9781 & 0.9886 & 9.833 & 37.99 & 36.92 & 0.9689 & 0.9849 & 9.815 \\
		
		 \scriptsize{GANHDR}\cite{HDR_GAN_2021} & 38.90 & 38.95 & \textbf{\textcolor{black}{0.9887}} & 0.9931 & 9.817 & 38.36 & 38.46 & \textbf{\textcolor{black}{0.9878}} & {\color{black} \underline{0.9922}} & 9.823 \\
		
		ADNet\cite{ADNet} & {\color{black} \underline{39.89}} & 40.50 & 0.9810 & 0.9903 & 9.880 & {\color{black} \underline{39.49}} & {\color{black} \underline{39.39}} & 0.9767 & 0.9897 & {\color{black} \underline{9.864}} \\
		\hline
		\textbf{SJ-HD$^2$R} & \textbf{\textcolor{black}{42.23}} & \textbf{\textcolor{black}{48.79}} & 0.9840 & \textbf{\textcolor{black}{0.9959}} & \textbf{\textcolor{black}{9.911}} & 
		\textbf{\textcolor{black}{42.71}} & \textbf{\textcolor{black}{48.08}} & 0.9824 & \textbf{\textcolor{black}{0.9994}}  & \textbf{\textcolor{black}{9.988}} \\
		\bottomrule[1pt]
	\end{tabularx}{}}
	\vspace{2mm}
	\caption{Quantitative comparison between the medium-exposure based baselines and our proposed method on RGB-HDR dataset and RAW-HDR dataset. The best and the second best performance are bold and underlined respectively.}
	\label{tab_comp_sota}
	\vspace{-2mm}
\end{table*}

\begin{table*}[!t]
	\begin{center}
	 \resizebox{0.71\textwidth}{!}{
	 \vspace{-4mm}
	\begin{tabular}{l | c c c c c c}
		\toprule[1pt]
		Method &  P-$\mu \uparrow$ & P-$L \uparrow$ & S-$\mu \uparrow$ & S-$L \uparrow$ & VDP3-$L$ $\uparrow$  & VDP3-$\mu$ $\uparrow$\\ 
		\hline
		
		Sen\cite{sen2012robust} & 35.76 & 39.75  & 0.9483  & 0.9438  & 9.6247  &9.4268\\
		
		Hu\cite{hu2013hdr} & 30.78 & 38.27 & 0.9548 & 0.9872 & 9.5828 & 9.1940\\
		
		Kalantari\cite{kalantari2017deep} & 32.58 & 48.22 & 0.9016  & 0.9988 & 9.9688 & 8.9687\\
		
		Wu\cite{wu2018deep} & 39.64  & 49.20 & 0.9705 & 0.9971 & 9.9654 & 9.6778\\
		
		SCHDR\cite{prabhakar2019fast} & 34.31 & 45.89 & 0.9124 & 0.9869 & 9.7974 & 9.2913\\
		
		AHDR\cite{yan2019attention} & 40.12 & 55.20 & 0.9732 & 0.9986 & 9.9832 & 9.7135\\
		
		DWT\cite{dai2021waveletbased} & 40.09 & 52.71 & 0.9751 & 0.9983  & 9.9752 & 9.7280\\
		
		GANHDR\cite{HDR_GAN_2021}  & 39.87  & 53.36  & \textbf{\color{black}{0.9862}}  & {\color{black} \underline{0.9993}} & 9.9720 & 9.6706\\
		
		ADNet\cite{ADNet}  & 41.99 & {\color{black} \underline{59.80}} & 0.9823  & 0.9896  & 9.9832 & 9.7858\\
		
		KPN\cite{Godard_2018_ECCV}  & 41.09 & 53.21 & 0.9742  & 0.9931  & 9.9809 & 9.7278\\
		
		
        
	    \hline			
        \textbf{SJ-HD$^2$R}($U_{ref}$) & \textbf{\color{black} {43.38}} & \textbf{\color{black}{61.56}} & {\color{black} \underline{0.9847}} & \textbf{\color{black} {0.9995}} & \textbf{\color{black} {9.9837}} & 9.8147 \\
		\textbf{SJ-HD$^2$R}($M_{ref}$) & 39.90 & 42.23 & 0.9805 & 0.9895 & 9.8311 & {\color{black} \underline{9.8584}} \\
		\textbf{SJ-HD$^2$R}($O_{ref}$) & 36.38 & 36.27 & 0.9736 & 0.9766 & 9.4260 & 9.7726 \\
		\textbf{SJ-HD$^2$R}($S_{ref}$) & {\color{black} \underline{42.23}} & 48.79 & 0.9840 & 0.9959 & {\color{black} \underline{9.9109}} & \textbf{\color{black} {9.8760}} \\
		\bottomrule[1pt]
	\end{tabular}}
	\end{center}
	\vspace{-1mm}
	\caption{Quantitative comparison with under-exposure based baselines, as well as deep burst denoising method on RGB-HDR dataset. In four variants of {SJ-HD$^2$R}, $U_{ref}$, $M_{ref}$, $O_{ref}$, $S_{ref}$ represent using under-exposure, medium-exposure, over-exposure, or selective exposure as reference respectively.}
	\label{tab_comp_ref}
	\vspace{-5mm}
\end{table*}

\begin{figure*}[t]
    \centering
    \includegraphics[width=0.97\textwidth, height=0.95\textwidth]{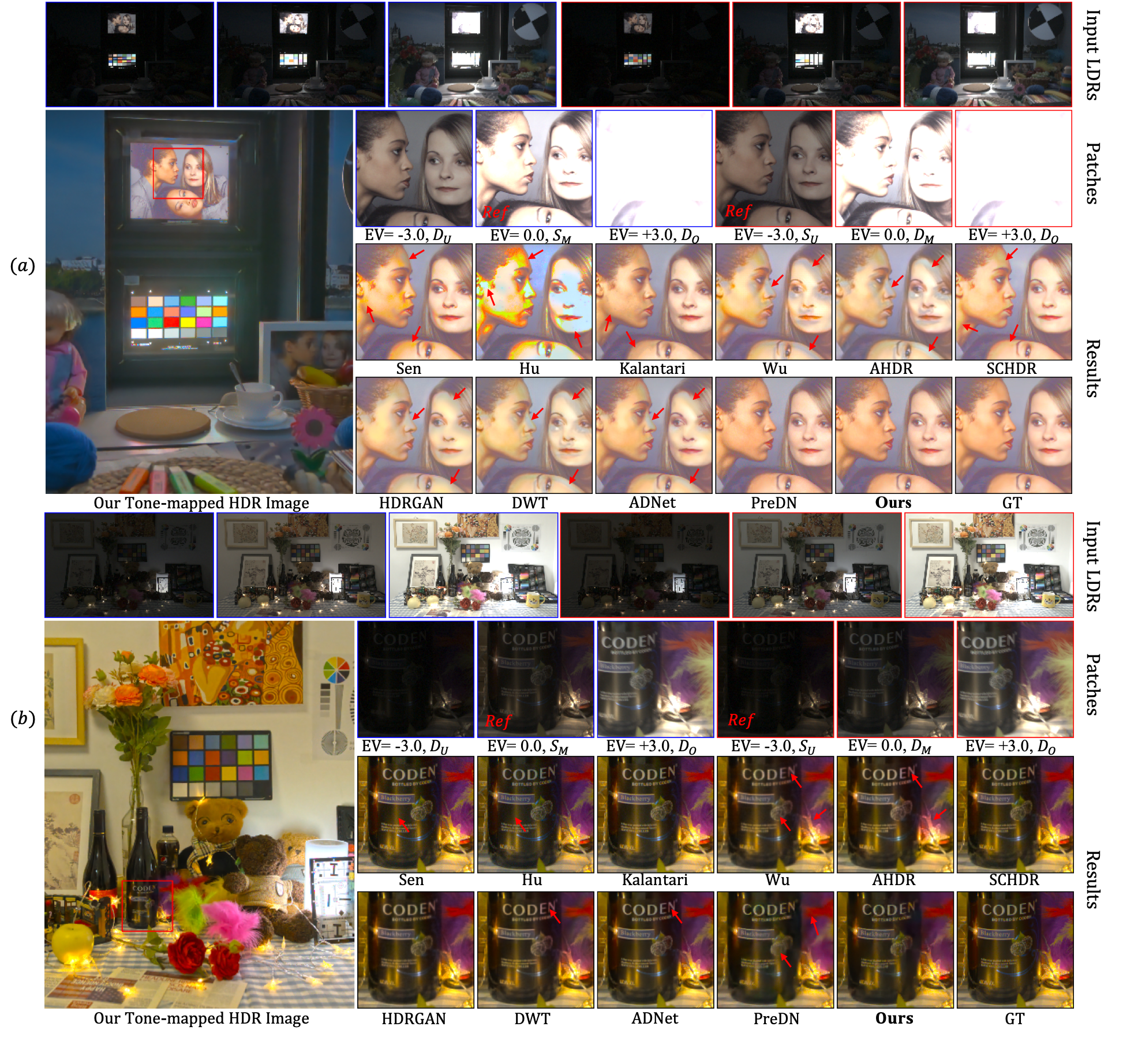}
    \vspace{-2mm}
    \caption{Qualitative comparison with previous works. Note that previous works use medium-exposure as the reference (the patches in blue rectangle), these two settings share the same ground truth. $(a)$ comparison in high-light region with occlusion. ${(b)}$ comparison in dark region with heavy noise.}
    \label{fig:figure_1}
    \vspace{-2mm}
\end{figure*}

\subsection{Evaluation Benchmark}
{\textbf{\emph{Datasets.}}}
Since there is no public HDR dataset that supports the switching of the reference, and no dataset has considered the noise removal in HDR scenes, so we create a new HDR benchmark that contains two datasets: the RAW-HDR dataset and RGB-HDR dataset, each of which consists of 192 pairs for training and 15 pairs for testing. We collect the datasets following the method in~\cite{kalantari2017deep}, but reserve all images and meta information to support using arbitrary LDR image as the reference. Note that these different settings, e.g., under-exposure based setting: $[S_U, D_M, D_O]$ and medium-exposure based setting: $[D_U, S_M, D_O]$, share the same ground truth, as the ground truth comes from the same static scene. Note that here $S$ and $D$ represent the image is from static scene and dynamic scene respectively. In addition, our datasets are of much more difficult dynamic scenes , either it is near-zero in brightness (varies from -0.5 to +8.7) or is corrupted by heavy noise ($ISO$ value varies from 50 to 3200). More details about the datasets can be found in supplementary material. We also use public datasets~\cite{kalantari2017deep,prabhakar2019fast} for qualitative validation.

{\textbf{\emph{Metrics.}}}
We use PSNR, SSIM and HDR-VDP-3 as the evaluation metrics. We compute the PSNR, and SSIM metrics both in tone-mapped domain using $\mu$-law (P-$\mu$, S-$\mu$) and linear domain (P-$L$, S-$L$, VDP3-$L$), which is consistent with previous works. Furthermore, we compute the HDR-VDP-3~\cite{HDR-VDP-3} (VDP3-$\mu$, VDP3-$L$) to evaluate the perceptual reconstruction quality with diagonal display size set to 24 inches, and the viewing distance set to 0.5 meter. 

\subsection{Comparison with State-of-the-arts}
For fair comparison, we retrain previous model-based baselines on our RGB and RAW datasets by running their public codes. We compare our model with all previous important works including 
medium-exposure based methods \cite{sen2012robust,hu2013hdr}, \cite{kalantari2017deep,wu2018deep,prabhakar2019fast,yan2019attention,dai2021waveletbased,HDR_GAN_2021,ADNet}, as well as burst denoising method\cite{Godard_2018_ECCV}.

\textbf{\emph{Comparison with methods using medium-exposure as reference.}} Table~\ref{tab_comp_sota} shows the quantitative comparison between our method and existing baselines on proposed RGB-HDR dataset and RAW-HDR dataset.
Note that all baselines use the medium-exposure as the reference frame. We run the public code for Sen~\cite{sen2012robust}, Hu~\cite{hu2013hdr} and Kalantari~\cite{kalantari2017deep}, and retrain models of Wu~\cite{wu2018deep}, SCHDR~\cite{prabhakar2019fast}, AHDR~\cite{yan2019attention}, DWT~\cite{dai2021waveletbased}, GANHDR~\cite{HDR_GAN_2021} and ADNet~\cite{ADNet} to obtain their best performance. As shown in Table~\ref{tab_comp_sota}, our proposed method outperforms all baselines on multiple metrics: PSNR-$\mu$, PSNR-$L$, SSIM-$L$ and HDR-VDP-3 by large margins, with a 2.3+ dB higher on PSNR-$\mu$ and a 7.9+ dB higher on PSNR-$L$ than the second best. (Note that ADNet~\cite{ADNet} is the winner of NTIRE2021 challenge on multiple-frames HDR fusion track~\cite{perez2021ntire}). We also compare our model with these baselines on Kalantari dataset\cite{kalantari2017deep} using the medium exposure as reference and our PCFNet achieve state-of-the-art performance quantitatively and qualitatively, as shown in Table~\ref{tab_comp_kalantari}.

\begin{table*}[h]
	\begin{center}
	 \resizebox{0.8\textwidth}{!}{
	\begin{tabular}{l | c c c c c}
		\toprule[1pt]
		Method &  PSNR-$\mu \uparrow$ & SSIM-$\mu \uparrow$ & PSNR-$L \uparrow$ & SSIM-$L \uparrow$ & HDR-VDP-2 $\uparrow$\\ 
		\hline
		
		Sen\cite{sen2012robust} & 40.9453 & 0.9085 & 38.3147  & 0.9749  & 60.5425\\
		
		Hu\cite{hu2013hdr} & 32.1872 & 0.9716 & 30.8395  & 0.9511  & 57.8278\\
		
		Kalantari\cite{kalantari2017deep} & 42.7423 & 0.9877 & 41.2518  & 0.9845  & 64.6519\\
		
		Wu\cite{wu2018deep} & 41.6377 & 0.9869 & 40.9082  & 0.9847  & 58.3739\\
		
		NHDRNet\cite{yan2020deep} & 42.4769 & 0.9942 & 40.1978  & 0.9889  & 63.1585\\
		
		SCHDR\cite{prabhakar2019fast} & 40.4700 & 0.9931 & 39.6800  & 0.9899  & 63.6192\\
		
		AHDR\cite{yan2019attention} & 43.6172 & {\color{black} \underline{0.9956}} & 41.0309  & 0.9903  & 64.8465\\
		
		DWT\cite{dai2021waveletbased} & 43.6734 & {\color{black} \underline{0.9956}} & 41.2195  & {\color{black} \underline{0.9905}}  & 64.9472\\
		
		HDR-GAN\cite{HDR_GAN_2021} & {\color{black} \underline{43.9220}} & 0.9865 & {\color{black} \underline{41.5720}}  & {\color{black} \underline{0.9905}}  & \textbf{\color{black} {65.4500}}\\
		
		DAHDR~\cite{yan2022dual} & 43.8400 & {\color{black} \underline{0.9956}} & 41.3100  & {\color{black} \underline{0.9905}}  & 64.6765\\
		\hline
		\textbf{PCFNet(Ours)} & \textbf{\color{black} {43.9372}} & \textbf{\color{black} {0.9959}} & \textbf{\color{black} {41.5784}}  & \textbf{\color{black} {0.9916}}  & {\color{black} \underline{65.3332}}\\
		\bottomrule[1pt]
	\end{tabular}}
	\end{center}
	\caption{Quantitative comparison on dataset~\cite{kalantari2017deep}. The best and the second best performance are bold and underlined respectively.}
	\label{tab_comp_kalantari}
	\vspace{-5mm}
\end{table*}

This result is reasonable and explainable. As discussed in previous section, we selectively decide the reference with high accuracy more than 93.3$\%$, so that we take advantages of ghost-free and fine details in highlight and high PSNR at the same time. This can be reflected on PSNR-$L$ and SSIM-${L}$, which may be dominated by the highlight luminance in a image. For metrics in tone-mapped domain, it still achieves the SOTA performance by large margins on PSNR-$\mu$ and comparable figures on SSIM-${\mu}$, which further suggests the effectiveness of our proposed solution. Same conclusions can be drawn on RAW-HDR dataset. Since \cite{kalantari2017deep,prabhakar2019fast} do not support calculating the optical flow for RAW image, we leave this part with "-" in Table~\ref{tab_comp_sota}. 

We also show several examples of qualitative comparison in Figure~\ref{fig:figure_1}. All results are displayed after tone-mapping by $Photoshop Software$~\cite{photoshop} using the same adjustment setting. The input LDR patches with different exposure as the reference are showed in the top row of each figure. We reiterate that these two sets share the same ground truth, so that their performance can be compared. As shown in Figure~\ref{fig:figure_1} $(a)$, in the case of ultra-high dynamic range scene that occlusion occurs in saturated regions, our proposed method obtain a ghost-free results compared with other baselines that suffer from missing details and ghosting artifacts. In addition, we also show another scene that has heavy noise in under-exposed regions. As shown in Figure~\ref{fig:figure_1} $(b)$, our proposed method have better or comparable quality in dark regions compared with medium-exposed based methods~\cite{sen2012robust,hu2013hdr,kalantari2017deep,wu2018deep,yan2019attention,prabhakar2019fast,HDR_GAN_2021,dai2021waveletbased,ADNet}.  

\textbf{\emph{Comparison with methods using under-exposure as reference.}} To further validate the contribution of our proposed framework, we compare our model with all baselines that use the under-exposed frame as the reference. Apart from~\cite{sen2012robust,hu2013hdr,kalantari2017deep}, we retrain all models involved on RGB-HDR dataset. As shown in Table~\ref{tab_comp_ref}, our model outperforms all previous baselines by large margins. We also provide qualitative comparison in dark regions in the supplement. 

\begin{figure}[h]
	\centering
	\includegraphics[width=0.67\columnwidth]{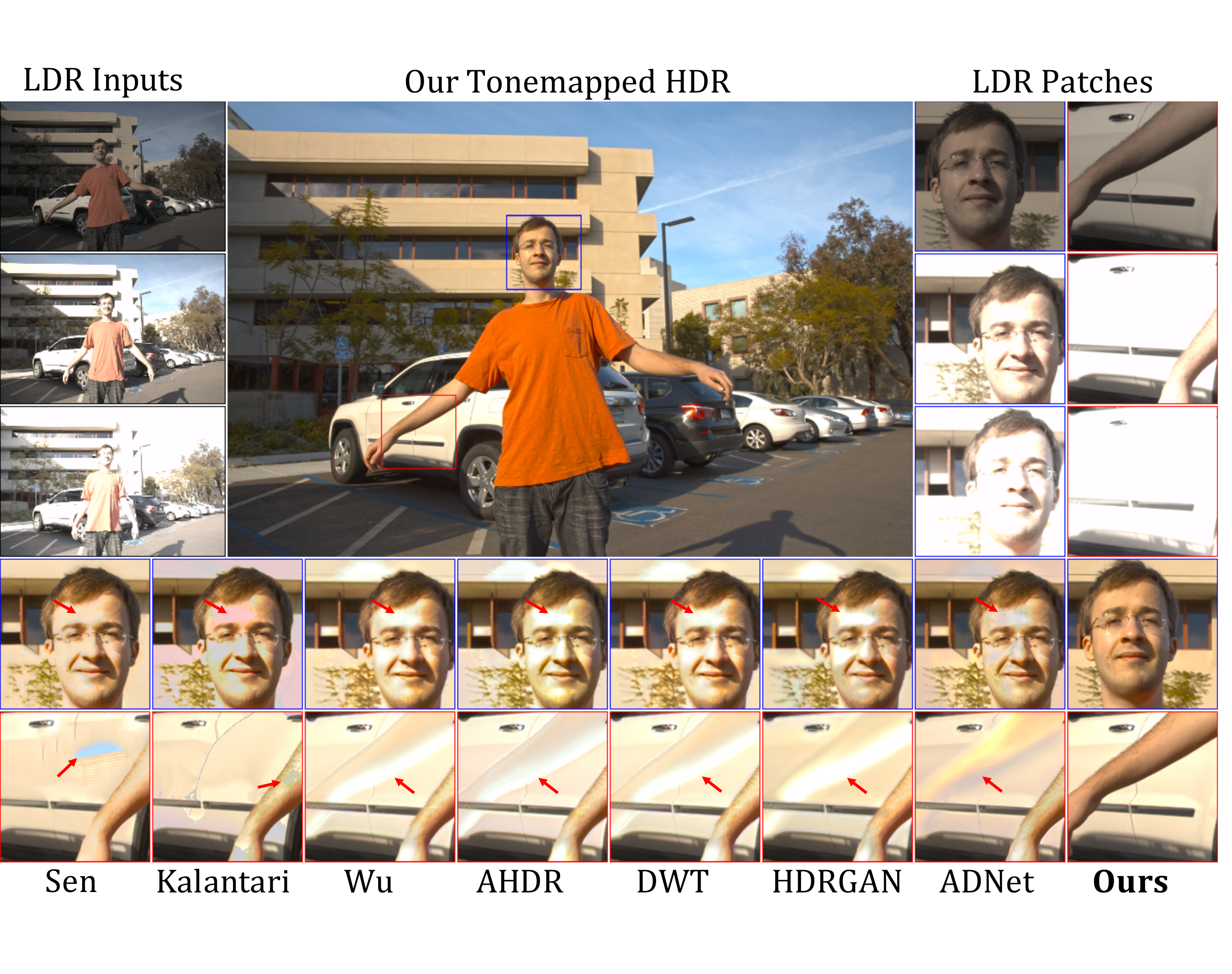}
	\vspace{-2mm}
	\caption{A comparison with best seven baselines in dynamic scene from dataset~\cite{kalantari2017deep}. Note that all methods take as input the same three LDR images.}
	\label{fig:figure_3}
	\vspace{-3mm}
\end{figure}

\textbf{\emph{Comparison on public datasets without GT.}} We also run the test on several public datasets: Kalantari~\cite{kalantari2017deep} dataset, Prabhakar~\cite{prabhakar2019fast} and Samsung dataset~\cite{Hu_2020_CVPR_Workshops}. Because of all these datasets discard the medium-exposed image of dynamic scenes, we can not compare them fairly using the same setting as we did in Table~\ref{tab_comp_sota}. Instead, we give a qualitative comparison that take the same three LDR images as input. So that our model and other baselines may generate an HDR image with different foreground content due to the different choice of reference. As shown in Figure~\ref{fig:figure_3}, that our result is ghost-free, no color distortion, and is of faithful details in high-light regions and is of comparable details with baselines in dark regions.

\subsection{Ablation Studies}
\textbf{\emph{The switching of reference frame.}} {To explore the performance of strategies using different frame as reference and our selective method, we compare four variants using different reference settings:}
\subitem{\textbf{{SJ-HD$^2$R}}($U_{ref}$)}: our model using under-exposure as reference image.
\subitem{\textbf{{SJ-HD$^2$R}}($M_{ref}$)}: our model using medium-exposure as reference image.
\subitem{\textbf{{SJ-HD$^2$R}}($O_{ref}$)}: our model using over-exposure as reference image.
\subitem{\textbf{{SJ-HD$^2$R}}($S_{ref}$)}: our model using selective exposure as reference image.

\begin{figure}[t]
	\centering
	\includegraphics[width=0.87\columnwidth]{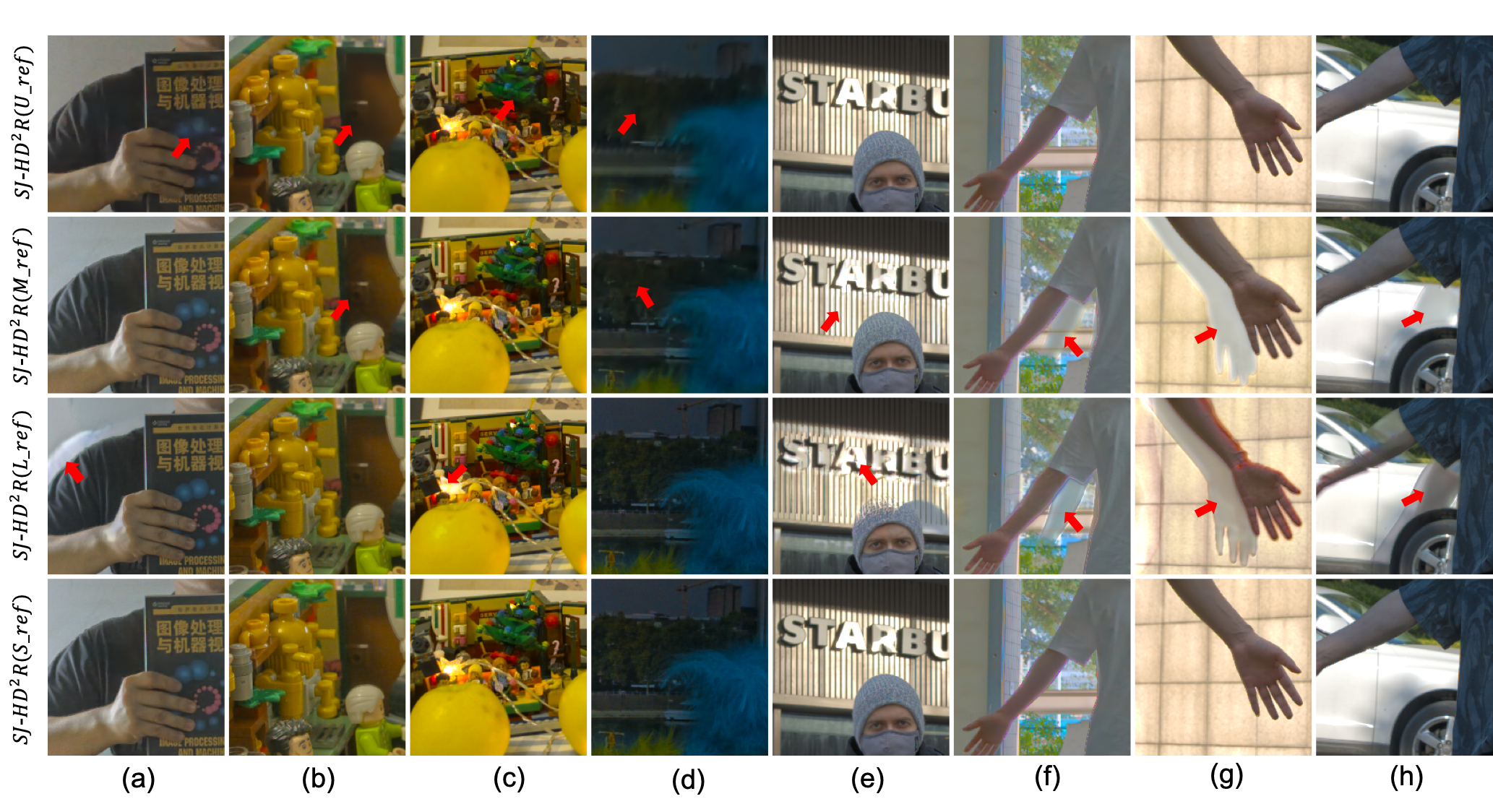}
	\vspace{-3mm}
	\caption{Examples of our selective strategy from our RGB-HDR dataset. The top three rows display the result of {SJ-HD$^2$R}($U_{ref}$), {SJ-HD$^2$R}($M_{ref}$), {SJ-HD$^2$R}($O_{ref}$), and the bottom row shows our selective results: {SJ-HD$^2$R}($S_{ref}$).}
	\label{fig:figure_2}
	\vspace{-2mm}
\end{figure}

From the results in Table~\ref{tab_comp_ref}, {SJ-HD$^2$R($U_{ref}$)} outperforms others on multiple metrics: P-$\mu$, P-$L$, S-$L$. {SJ-HD$^2$R($M_{ref}$)} outperforms {SJ-HD$^2$R($M_{ref}$)} on S-$\mu$ because it has richer details in the dark. {SJ-HD$^2$R($S_{ref}$)} outperforms others on VDP3-$\mu$, as we optimize the path selection using metrics VDP3-$\mu$. In Figure~\ref{fig:figure_2} we show the qualitative results of path selection, which can further prove our method's effectiveness.
\vspace{1mm}

\textbf{\emph{Modules in our SJ-HD$^2$R framework.}} {We investigate the variants of our framework and validate the contribution of each component in SJ-HD$^2$R. We conduct this ablation by comparing several variants: (i) SJ-HD$^2$R(w/o TMO): without TMO and InvTMO and use $\mu$-law to calculate tone-mapped loss; (ii) SJ-HD$^2$R(w/o MSA): without multi-scale attention modules in PCFNet; (iii) SJ-HD$^2$R: the full version of our method. Table \ref{tab_ablation_1} suggests that compared with SJ-HD$^2$R the TMO/InvTMO and the MSA contributes +1.78 dB and +0.88 dB to SJ-HD$^2$R respectively.}
\vspace{1mm}

\textbf{\emph{Pre-denoising network.}} {We do the ablation study on PreDNNet by comparing SJ-HD$^2$R with several variants: (i) PreDN: our pre-trained denoising network only. (ii) SJ-HD$^2$R(w/o PreDN): without pre-denoising, taking noisy LDR reference image as input only. (iii) ADNet(U): the best baseline ADNet~\cite{ADNet} with denoised under-exposure by PreDNNet. (iv) SJ-HD$^2$R(U): SJ-HD$^2$R with denoised under-exposure by PreDNNet. (v) SJ-HD$^2$R(U+M): SJ-HD$^2$R with denoised under and medium exposures. (vi) SJ-HD$^2$R: the full version of our method with denoised three exposures. Table \ref{tab_ablation_1} suggests that compared with SJ-HD$^2$R: (i) the more LDRs are processed by PreDNNet, the better performance can be obtained. (ii) PreDNNet only can not reconstruct faithful HDR image. (iii) Our SJ-HD$^2$R outperforms the best previous HDR method with same pre-denoising processing.}

\textbf{\emph{Tone mapping operators.}} {To explore the best TMO that converts an HDR image to a histogram-balanced tone-mapped domain, we try most widely used TMO methods, including ${Linear}$ TMO that represents HDR image itself, the ${Gamma}$ curve with $\gamma$=2.2, the ${Reinhard}$~\cite{ReinhardTMO} TMO, the ${Heble}$~\cite{hable2010uncharted} TMO, and $\mu$-law TMO. The formulations of these TMOs and their corresponding InvTMO can be found in supplementary material. Table~\ref{tab_ablation_TMO} shows the performance of different kinds of TMO, from which we can find that the ${\mu}$-law with $\mu$=5000 outperforms other kinds of TMOs significantly.}\par

\begin{table}[tb] \small
	\begin{center}
	\resizebox{0.6\columnwidth}{!}{
		\begin{tabular}{l | c c c c}
			\toprule[1pt]
			Method & PSNR-$\mu$ & PSNR-$L$ & SSIM-$\mu$ & SSIM-$L$\\
			\hline
			SJ-HD$^2$R(w/o TMO) & 41.60 & 58.18 & 0.9808 & 0.9992 \\
			SJ-HD$^2$R(w/o MSA) & 42.50 & 60.21 & 0.9829 & 0.9994 \\
	    	\hline
	    	PreDN & 41.38 & 59.68 & 0.9799 & 0.9992 \\
	    	SJ-HD$^2$R(w/o PreDN) & 42.95 & 60.46 & 0.9839 & 0.9995\\
	    	\small{ADNet(U)} \cite{ADNet}  & 42.08 & 60.59 & 0.9825 & 0.9992 \\
	    	SJ-HD$^2$R(U) & 43.21 & 60.64 & 0.9842 & 0.9995 \\
	    	SJ-HD$^2$R(U+M) & 43.37 & 61.54 & 0.9846 &  \textbf{0.9996} \\
	    	SJ-HD$^2$R & \textbf{43.38} & \textbf{61.56} & \textbf{0.9847} & 0.9995 \\
			\bottomrule[1pt]
		\end{tabular}}
	\end{center}
	\caption{Ablation study on variants of our framework. We conduct these ablation on SJ-HD$^2$R$(U_{ref})$ and all scores are the average across all testing images.}
	\label{tab_ablation_1}
	\vspace{-8mm}
\end{table}

\begin{table}[tb] \small
	\begin{center}
	\resizebox{0.5\columnwidth}{!}{
		\begin{tabular}{l | c c c c}
			\toprule[1pt]
			Method & PSNR-$\mu$ & PSNR-$L$ & SSIM-$\mu$ & SSIM-$L$\\
			\hline
			Linear & 39.56 & 54.28 & 0.9726 & 0.9987 \\
			Gamma2.2 & 42.44 & 60.66 & 0.9821 & 0.9994 \\
			Reinhard~\cite{ReinhardTMO} & 41.56 & 58.82 & 0.9796 & 0.9992 \\
			Hable~\cite{hable2010uncharted} & 41.20 & 57.66 & 0.9782 & 0.9991 \\
			${\mu}$-Law & \textbf{43.21} & \textbf{60.40} & \textbf{0.9842} & \textbf{0.9995} \\
			\bottomrule[1pt]
		\end{tabular}}
	\end{center}
	\caption{Ablations on TMOs. We conduct the ablation on path  SJ-HD$^2$R$(U_{ref})$.}
	\label{tab_ablation_TMO}
	\vspace{-8mm}
\end{table}


\section{Limitation and Discussion}
Admittedly, our models have limitations. A perfect compromise on exposure time for under or medium exposure reference sometime does not exist because of the ultra-high dynamic range of the real world. Though we have had great alleviation for each model and can adaptively predict the best model from scenario priors, ghost artifacts or degradation may still exist either in under-exposure based model or medium-exposed model. Additionally, all models in Figure~\ref{fig:overall_structure} are pre-trained and weight-freeze during inference, this increases the complexity and time of the training progress. To make the end-to-end optimizing feasible, we can seek soft-selection implementation or borrow experience from reinforcement learning~\cite{path_restore2021,rl_survey}, utilizing current metrics as its reward in the future work.
\section{Conclusion}
In this paper, we proposed a novel joint HDR and denoising imaging framework for dynamic scenes. We develop multiple efficient HDR fusion models that support using different exposure as reference, aiming to solve various types of side effects. Then we propose an selective inference strategy by utilizing image and scenario priors.
Thanks to these novel findings, our proposed method demonstrates superior performance to previous baselines.\par



%
%
\bibliographystyle{splncs04}
\bibliography{egbib}
\end{document}